\documentclass[sinclecolumn]{aastex631}

\submitjournal{AAS Research Notes}

\shorttitle{PGIR20dci}
\shortauthors{Hodapp et al.}
\begin{document}
\turnoffedit
\title{The historic $K_s$ light curve of the FUor PGIR20dci}

\correspondingauthor{Klaus Hodapp}
\email{hodapp@ifa.hawaii.edu}

% \author[0000-0003-0786-2140]{Klaus W. Hodapp}
\author{Klaus W. Hodapp}
\affil{University of Hawaii, Institute for Astronomy, 640 N. A'ohoku Place, Hilo, HI 96720, USA}

\author{Scott E. Dahm}
\affil{U.S. Naval Observatory, Flagstaff Station, 10391 West Naval Observatory Road, Flagstaff, AZ 86005-8521, USA}

\author{Watson P. Varricatt}
\affil{University of Hawaii, Institute for Astronomy, 640 N. A'ohoku Place, Hilo, HI 96720, USA}



\begin{abstract}

We report a historic $K_s$-band light curve spanning over three decades
of the FUor PGIR20dci recently discovered by
\citet{Hillenbrand.2021.AJ.161.220H.PGIR20dci} .
We find some minor variability of the object prior to the
FUor outburst, an initial rather slow rise in brightness, followed in 2019 by a much steeper rise to the maximum.

\end{abstract}




\section{Introduction} \label{sec:intro}

The FUor eruption of the Class I object PGIR20dci in the NGC281W star-forming region
was recently discovered by \citet{Hillenbrand.2021.AJ.161.220H.PGIR20dci} in the Palomar Gattini Infrared (PGIR) survey.
Their near-infrared spectroscopy confirmed that this
object is currently undergoing a FUor \citep{Herbig.1977.ApJ.217.693H}
outburst.
The published light curve relied on Spitzer, WISE and NEOWISE
photometry, with WISE having rather poor spatial resolution. 
This Research Note reports the pre-outburst and outburst
light curve based on ground-based $K_s$-band images. \\

% Fig. 1
\begin{figure}[h]
\begin{center}

\includegraphics[angle=0.,scale=1.0]{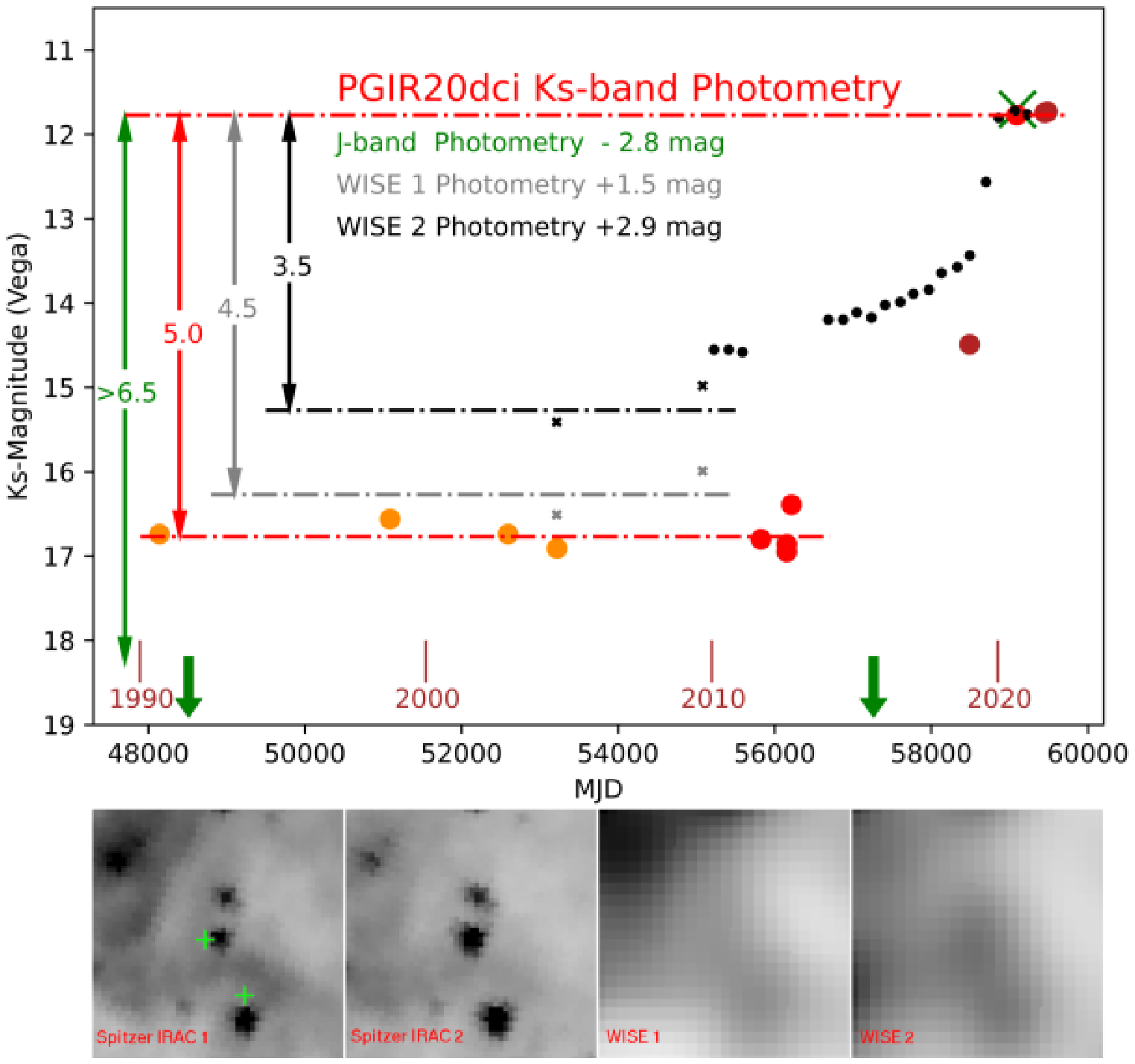}
\caption{ \\
Ks-band light curve of PGIR20dci based on ground-based observations.
All observations have been transformed into effectively the $K_s$ band.
The color of the symbols indicates the original filter of the observations:
$K^\prime$ (orange), $K_s$ (red), and $K$ (dark red). For comparison, we also show the WISE and NEOWISE
band 2 data points (black circles), as well as the Spitzer space telescope data in
the IRAC bands 1 (grey X) and 2 (black X) bands
reported by \citet{Hillenbrand.2021.AJ.161.220H.PGIR20dci} after color transformation into
the WISE photometric system. The flux of a constant star $\approx$ 5$\arcsec$ to the north
has been subtracted from the WISE-2 data points.
We show the $J$-band upper limit, the average plateau brightness reported by
 \citet{Hillenbrand.2021.AJ.161.220H.PGIR20dci} in green and also include their
$K_s$ data point in red.
All magnitudes are
shifted to match the outburst plateau $K_s$ level.
The photometric errors are $\approx$0.1 mag, about the size of the filled circles
used to indicate the data.
The outburst amplitudes in all the bands shown here are indicated by the arrows
on the left side of the Figure.
The 30$\arcsec\times$30$\arcsec$ images are Spitzer IRAC images from 2009 Sept. 3 and WISE images taken
between Jan. 2010 and Feb. 2011. The WISE catalog positions are indicated by green crosses.
 }
\end{center}
\end{figure}

\section{Observations and Results}

We have extracted aperture photometry from K-band images
of PGIR20dci in our data archive, using the same 2$\farcs$25 aperture radius
as was used by  \citet{Hillenbrand.2021.AJ.161.220H.PGIR20dci} for their $K_s$ data
point.
The first image was obtained in 1990 as part of the $K^\prime$ filter 
imaging survey
of star-forming regions by \citet{Hodapp.1994.ApJS.94.615H.KpSurvey}. 
Follow-up imaging data were obtained in $K^\prime$ using the QUIRC
\citep{Hodapp.1996.NewA.177H.HAWAIIdet}
camera on the UH 2.2m telescope in 1998, 2002, and 2004.
Later, three sets of images were obtained using WIRCam
\citep{Puget.2004.SPIE.5492.978P.WIRCAM}  on the CFHT
telescope in 2011 and 2012 using a  $K_s$ filter.
PGIR20dci was recorded with WFCAM \citep{Casali.2007.WFCAM} on UKIRT 
as part of the US Naval Observatory UKIRT Northern Hemisphere
$K$-band survey UHSK
\citep{Dahm.2018.AAS.23143604D-USNOKsurvey} in 2019.
Finally, two outburst images in $K$ were obtained
with UKIRT WFCAM in 2021.
The photometric calibration is based on a set of 25 2MASS \citep{Skrutskie.2006.AJ.131.1163}
stars 
with perfect 2MASS photometry flags
common to all the $K$-band images used here.
Based on the
color information near maximum
by \citet{Hillenbrand.2021.AJ.161.220H.PGIR20dci} and interpolation of
the color corrections given by \citep{Wainscoat.1992.AJ.103.332W.KpFilter}
we correct the early $K^\prime$ photometry towards brighter
values by subtracting 0.08 mag, and we correct the WFCAM photometry
by adding 0.12 mag, so that the data shown in Fig.~1 and listed in Table 1 
are effectively in the 
$K_s$ system. 
For clarity, we list the filter used in each observation in Table 1.
We include two upper limits at $J$ = 21 mag based on the non-detection
of the object in the
UKIRT Hemisphere Survey in $J$ (UHSJ) \citep{Dye.2018.MNRAS.473.5113.UHSJ} 
and the image published by
\citet{Megeath.1997.AJ.114.1106.NGC281W}.
In either image, only a faint extended ''smudge'' is visible at the position
of the reflection nebula, as was noted by  \citet{Hillenbrand.2021.AJ.161.220H.PGIR20dci}.
We show the $J$ brightness at the plateau level from \citet{Hillenbrand.2021.AJ.161.220H.PGIR20dci}
as a single datum (green X) at $J$ = 14.5 (but shifted in Fig. 1).\\

\begin{deluxetable}{rrrl}
\tabletypesize{\scriptsize}
\tablecaption{$K_s$ Photometry of PGIR20dci}
\tablewidth{0pt}
\tablehead{
\colhead{MJD} & \colhead{K$_s$} & \colhead{Error} & \colhead{Orig. Filter}
}
\startdata
48142 & 16.74 & 0.10 & $K^\prime$ \\
51085 & 16.56 & 0.10 & $K^\prime$ \\
52592 & 16.74 & 0.10 & $K^\prime$ \\
53218 & 16.91 & 0.10 & $K^\prime$ \\
55823 & 16.80 & 0.10 & $K_s$ \\
56152 & 16.95 & 0.10 & $K_s$ \\
56152 & 16.86 & 0.10 & $K_s$ \\
56213 & 16.39 & 0.10 & $K_s$ \\
58487 & 14.49 & 0.10 & $K$ \\
59089 & 11.77 & 0.10 & $K_s$ \\
59447 & 11.74 & 0.10 & $K$ \\
59481 & 11.73 & 0.10 & $K$ \\
\enddata
\end{deluxetable}

\newpage
 
% Section 4.
\section {Discussion}

The UH 2.2m and CFHT data between MJD 48000 ($\approx$ 1990) and 56500 ($\approx$ 2011) show only minor
variations in the range of approximately 0.5 mag, which we consider
significant compared to the 0.1 mag errors.
The Spitzer and WISE photometry presented by \citet{Hillenbrand.2021.AJ.161.220H.PGIR20dci}
had indicated a step increase of $\approx$ 2 mag in band 1 at wavelengths of 3.4 $\mu$m,
and a similar, though smaller, step in WISE band 2 at 4.5 $\mu$m, between MJD 55080 (2009) and 55222 (2010). 
Our $K_s$-band data do not show a similar step up in brightness between 
the UH 2.2m photometry prior to MJD 54000 and the 
CFHT photometry around MJD 56000.
We note that the WISE catalog profile-fitting positions and photometry are affected by the blending
of the signal of the FUor with extended emission
and another star's flux
$\approx$ 5$\arcsec$ north of PGIR20dci seen in the Spitzer images in Fig. 1.
Under the assumption that the flux ratio of that star to the FUor measured on the Spitzer image from 2009 also represents
their flux ratio in the blended first WISE data point from 2010, we
subtract the star's flux from this and all the other WISE measurements resulting in the light curve indicated by black dots
in Fig. 1 that shows a steady increase in brightness over the decade of the 2010s.
This technique works quite well, though not perfectly, for WISE band 2, where only a $\approx$ 0.4 mag
discontinuity between the last Spitzer and first WISE data point remains
that can be explained by the extended emission near PGIR20dci visible in the Spitzer band 1 and 2
images in Fig. 1.
This technique of subtracting the neighboring star's flux does not work equally well for Spitzer band 1, where PGIR20dci is
fainter than that constant star and the nebulous background is stronger.
The variations of a few tenths of magnitudes in the $K_s$ band, as well as the difference between the
two Spitzer epochs indicate minor variations of PGIR20dci prior to the onset of the steady brightening
sometime around 2014.
Both the $K_s$ data point at MJD 58487 (2019) and coincident NEOWISE data points are roughly half-way, in magnitudes,
between the quiescent levels and the outburst plateau level. After this time, a much more rapid
rise to the light curve maximum is indicated by two NEOWISE data points prior to the discovery
of the outburst by \citet{Hillenbrand.2021.AJ.161.220H.PGIR20dci}.
The outburst amplitudes measured in $K_s$ and in the WISE photometric system
are indicated by arrows in Fig. 1. 
They show the familiar trend of smaller
amplitudes at longer wavelengths.
The outburst of PGIR20dci reached a plateau around MJD 58900 (February 2020) and has held this
level at least until September of 2021.

\section{Summary and Conclusions}

We have presented a historic light curve of the recently discovered FUor outburst PGIR20dci.
The minor variations prior to the outburst are typical of T Tauri stars, the
precursors of most FUor outbursts.
The step increase in brightness reported by \citet{Hillenbrand.2021.AJ.161.220H.PGIR20dci}
is not confirmed by our $K_s$-band light curve, but
can be explained by blending with the flux of another nearby star and extended emission.
The transition from a slow to a rapid rise in brightness halfway to the maximum
is remarkable. The outburst amplitude is decreasing with wavelength.
PGIR20dci has now reached a stable maximum brightness.

\begin{acknowledgments}
This work uses imaging data from the WFCAM at the UKIRT observatory and from the NICMOS3 and QUIRC instruments at the UH 88" telescope, both operated by the University of Hawaii.
One data point is from the UHS-K survey done at UKIRT with funding from the U.S. Naval Observatory.
Some data were obtained with WIRCam, a joint project of CFHT, Taiwan, Korea, Canada, France, at the Canada-France-Hawaii Telescope (CFHT). 
The photometry is based on data products from the Two Micron All Sky Survey.
We thank Mike Irwin of the Cambridge Astronomical Survey Unit at University of Cambridge for his work in 
reducing the UKIRT data, and Lynne Hillenbrand for helpful discussions.\\
\end{acknowledgments}

\vspace{5mm}
\facilities{UH:2.2m, CFHT, UKIRT, WISE, Spitzer}



\end{document}